%% file: 1_MAIN.tex
\documentclass[runningheads]{llncs}

\usepackage[T1]{fontenc}
\usepackage{xcolor}
\usepackage[colorlinks=true, urlcolor=blue, linkcolor=blue, citecolor=blue]{hyperref}
\usepackage{graphicx}
\begin{document}

\makeatletter
\def\ps@firstpage{
  \let\@oddfoot\@empty
  \let\@evenfoot\@empty
  \def\@oddhead{\hfill Accepted to 27th International Conference on AI in Education (AIED) 2026 \hfill}
  \let\@evenhead\@oddhead
}
\makeatother

\title{Exploring Teachers' Perspectives on Using Conversational AI Agents for Group Collaboration}
\titlerunning{Teachers' Perspectives on Conversational AI Agents for Collaboration}
%
%
\author{Prerna Ravi\inst{1}\orcidID{0000-0002-4289-5610} \and
 Carúmey Stevens\inst{1}\orcidID{0009-0002-0326-5058} \and Beatriz Flamia Azevedo\inst{2}\orcidID{0000-0002-8527-7409} \and Jasmine David\inst{1}\orcidID{0009-0004-8791-8048} \and Brandon Hanks\inst{1}\orcidID{0009-0003-4514-362X} \and Hal Abelson\inst{1}\orcidID{0000-0002-5328-7821} \and Grace Lin\inst{1}\orcidID{0000-0001-7552-2880} \and Emma Anderson\inst{1}\orcidID{0000-0002-6561-9977}}

\institute{Massachusetts Institute of Technology, Cambridge MA, USA 
\\
\email{\{prernar,carumeys,bhanks,hal,gcl,eanderso\}@mit.edu, jas.dav7654@gmail.com}
\and Instituto Politécnico de Bragança, Bragança, Portugal
\\
\email{beatrizflamia@ipb.pt}
}

\authorrunning{P. Ravi et al.}
%
%
\maketitle              
\thispagestyle{firstpage}
\begin{abstract}
Collaboration is a cornerstone of 21st-century learning, yet teachers continue to face challenges in supporting productive peer interaction. Emerging generative AI tools offer new possibilities for scaffolding collaboration, but their role in mediating in-person group work remains underexplored—especially from the perspective of educators. This paper presents findings from an exploratory qualitative study with 33 K-12 teachers who interacted with Phoenix, a voice-based conversational agent designed to function as a near-peer in face-to-face group collaboration. Drawing on playtesting sessions, surveys, and focus groups, we examine how teachers perceived the agent’s behavior, its influence on group dynamics, and its classroom potential. While many appreciated Phoenix’s capacity to stimulate engagement, they also expressed concerns around autonomy, trust, anthropomorphism, and pedagogical alignment. We contribute empirical insights into teachers' mental models of AI, reveal core design tensions, and outline considerations for group-facing AI agents that support meaningful, collaborative learning.

\end{abstract}
\section{Introduction}

Collaboration is a fundamental skill of 21st-century education, recognized for its role in promoting communication, critical thinking, and social-emotional development \cite{liu2024}. 
However, fostering meaningful collaboration in classrooms remains a persistent challenge for teachers due to uneven participation, group conflict, and difficulties in assessing individual contributions \cite{Roberts2007}.

To address these challenges, researchers have explored various methods, including optimizing group formation \cite{tan2022systematic}, automated feedback to improve discourse \cite{dyke2013enhancing}, and tutor-facing tools to guide pedagogical interventions \cite{casamayor2009intelligent}. 
The rise of generative AI (GenAI) also introduces new possibilities for augmenting collaborative learning. 
Large language models (LLMs) can offer tailored dialogic feedback \cite{lu2025optimizing}, while playing dynamic roles in educational settings from assistants to collaborators \cite{Zhang2025}, but key gaps remain.
First, there is limited research on LLM agents in face-to-face collaborative learning settings, despite this being the predominant mode of classroom interaction \cite{tan2022systematic}. Second, few studies have explored agents designed as near-peers, rather than assistants or tutors, and investigated how their personas shape group dynamics, despite their known benefits in fostering equitable contributions, stimulating creative thinking, and supplementing knowledge gaps \cite{liu2024,lyu2026designing,perez2021review}. Finally, while research has primarily focused on students, there is a notable absence of studies on teachers’ perspectives on deploying peer conversational agents (CAs) in group work, even though teachers ultimately decide whether and how such tools are adopted in classrooms.

In this paper, we present findings from an exploratory qualitative study with 33 K-12 educators who interacted with Phoenix, a voice-based AI agent designed to act as a near-peer in group activities. 
We examine (RQ1) how teachers perceived Phoenix’s role in group discussions and its influence on collaboration, and (RQ2) how they envisioned integrating it into their classrooms. While teachers appreciated Phoenix’s potential to spark engagement and facilitate group talk, they also raised concerns about agency, trust, and pedagogical fit. We contribute insights into teachers’ mental models of group-facing AI, highlight tensions, and offer design implications for future collaborative learning agents.

\section{Related Work}
\input{2_RelatedWork}

\section{Agent Implementation}
\input{3_System}

\section{Methods}
\input{4_Methods}

\section{Results}
\input{5_Results}

\section{Discussion}
\input{6_Discussion}

%
%
%
\bibliographystyle{splncs04}
\bibliography{MAIN}
%





\end{document}

%% file: 2_RelatedWork.tex
\subsubsection{Group Collaboration in Education}
\label{lit_collab}

Collaboration in education is defined as a process in which two or more students work interdependently toward a shared goal, pooling their knowledge, skills, and efforts to produce a more effective outcome than possible individually \cite{evans2020measuring}. 
Well-structured collaborative learning promotes soft skills that support adaptability, including teamwork, communication, negotiation, and socio-emotional development \cite{hernandez2019computer}.
However, implementing classroom collaboration comes with challenges including unequal participation from free-riding, skill gaps, group selection and composition concerns, and difficulties with fair assessment \cite{Roberts2007}. 
To address these, a substantial body of work has focused on optimizing group composition \cite{tan2022systematic}, conversational agents (CAs) to improve interaction patterns \cite{dyke2013enhancing}, and instructor-facing recommendation systems for pedagogical interventions during group work \cite{casamayor2009intelligent}. 
However, most such tools have been evaluated in online settings rather than face-to-face classrooms, which remain the dominant context in K-12 education \cite{tan2022systematic}. Recent advances in LLMs that enable more naturalistic, responsive dialogue \cite{liu2024} create new opportunities to study AI-facilitated peer interaction in in‑person collaborative settings.


\subsubsection{AI Tools to Support Group Collaboration}

There has been an increase in GenAI systems for multi-party collaboration. 
Notable examples enhance the quality of group conversations \cite{anderson2025exploring}, intelligently prompt peers at opportune moments \cite{de2025investigating}, and improve group decision-making \cite{Chiang2024EnhancingAI}. In these studies, LLM-based CAs are increasingly being integrated not just as tools but team members that intellectually contribute to task execution. 
LLM chatbots can enhance group awareness and performance \cite{Amiot2025}, and help reduce social anxiety or communication deadlocks during classroom discussions \cite{Zhang2025}.
\textit{Peer} agents in particular can regulate emotional feedback, promote equality, and challenge student thinking \cite{perez2021review}, thereby promoting creative collaboration in learning contexts \cite{liu2024}.
However, LLM CAs can foster cognitive dependency, constrain learner autonomy, perpetuate cultural biases, and suppress creativity~\cite{Zhang2025}.
These tensions highlight the importance of thoughtfully deploying AI CAs in educational settings, grounded in the lived experiences of those who mediate their use. While research on peer group agents has mainly focused on students, teachers—the key decision-makers shaping pedagogy and technology integration—remain overlooked \cite{zawacki2019systematic}. Our study explores teacher perspectives on LLM peer agents, highlighting both promise and limitations for classroom use, \textit{before} being introduced to students.

\subsubsection{Perception of AI Conversational Agents}

Designing effective AI agents for collaboration requires not only robust communicative capabilities but also a careful understanding of how people \textit{perceive, interpret, and respond} to these agents. 
Humans apply social rules to computers and view them as social actors when there is interdependence between them \cite{nass1994computers}. 
Learners attribute human-like characteristics to AI-based tutors or companions, which can boost initial engagement but also influence trust and learning outcomes \cite{ackermann2025physical}. 
Importantly, perceptions of the agent’s proactivity and conversational fluency shape whether participants accept it as a legitimate collaborator \cite{Zhang2023Investigating}.
Positive perceptions can decrease cognitive overload during interactions \cite{Amiot2025}. Mixed perceptions of an AI agent’s reliability and unclear role can undermine its acceptance in group learning \cite{edwards2025human}. 
Participants’ preconceptions may also influence judgments of etiquette and risk for human-like agents \cite{Leong2024}. 
In education, perceptions of AI as a group member remain underexplored.
Given that collaboration relies on trust and shared responsibility, our study examines how \textit{teachers} perceive the role and pedagogical relevance of conversational agents, critical for effective classroom integration.

%% file: 3_System.tex
\subsubsection{Agent Design and Peer Persona} 
\label{agent-design}
Our system is a voice-based, non-embodied conversational agent designed to participate in real-time group discussions through spoken dialogue with multiple human users. Grounded in prior theory, we position the agent as an equal peer that supports knowledge co-construction, emotional regulation, and shared responsibility, in contrast to the authoritative, expert-led roles typical of AI tutors \cite{perez2021review,lyu2026designing}. This distinction reflects a shift from learning \textit{from} computers to learning \textit{with} them \cite{perez2021review}. Prior work shows that AI-supported collaboration can outperform human-only collaboration \cite{sankaranarayanan2020agent}, particularly when the agent operates at a similar knowledge level that avoids reinforcing group power dynamics \cite{Zhang2024_verbal}. Rather than a productivity tool, the agent is designed to scaffold productive struggle and equitable, human-led participation without increasing teacher burden, and offers content-agnostic audio support. 

It was hence introduced to teachers as a 30-year-old adult working with them. The agent was intentionally named \textbf{Phoenix} for ease of speech recognition. We picked a gender-neutral voice to reduce the influence of gender bias on perceptions \cite{piercy2025gender}.
Although Phoenix was voice-based, it was intentionally non-embodied to avoid confounds from visual features known to shape social authority and presence \cite{ackermann2025physical}. This allowed us to isolate how Phoenix's verbal behavior, tone, and timing influenced perceptions of its value, assigned trust, role, and social status.

\subsubsection{LLM Prompt} Phoenix was prompted to act as a constructive peer by building on group ideas and advancing the current topic, while avoiding an overly didactic tone. To preserve conversational flow, it was instructed to speak succinctly (approx. 20 words) without monopolizing airtime. During moments of drift or confusion, it could ask brief clarification questions to sustain momentum. Phoenix was also designed to express opinions confidently yet flexibly, to simulate a thoughtful peer. The prompts (\href{https://tinyurl.com/AIEDPersonaAppendix}{Appendix A.1}) included given task instructions and domain context to support meaningful contributions \cite{sun2024building}. 


\subsubsection{Technical Implementation}


We implemented a multi-layered system architecture (\href{https://tinyurl.com/AIEDPersonaAppendix}{Appendix A.2}) built to support real-time, voice-based interaction with Phoenix. The backend integrates Google’s real-time speech-to-text API for transcription, Microsoft Azure’s text-to-speech service for voice synthesis, and OpenAI’s GPT-4.1-mini as the core language model. 
The system runs on a Django backend using PostgreSQL for storage, with WebSockets enabling real-time bidirectional communication across layers, and an SFTP server managing secure transcript and metadata logging.
Core system layers include:

\begin{enumerate}
    \item \textbf{User Layer}:
This audio-based user interface provides a visual display of the agent’s voice output alongside a live transcript of the conversation.
 
    \item \textbf{GPT Layer}:
 This layer interfaces with the OpenAI API, sending user utterances to the model and returning context-aware dialogic responses.
    \item \textbf{Response Coordination Layer}:
 Sitting between the user and GPT layers, this layer handles transformation of incoming media, transcript analysis, and response regulation. 
To simulate natural turn-taking, a response gatekeeping system was implemented with GPT-4.1-mini to determine when and how Phoenix should respond. This was iteratively tested over multiple sessions and months prior to the study. 
It analyzes the eight most recent dialogue lines to see if the latest utterance is being directed at Phoenix and grants or denies it permission to respond. This binary classifier helps prevent irrelevant interjections. Phoenix is also triggered to respond when addressed by name. Common filler words and verbal pauses are filtered out to avoid false triggers.
 
\end{enumerate}

%% file: 4_Methods.tex
\subsection{Participants and Recruitment}
This study received an Institutional Review Board exemption from the authors’ institution. We recruited 33 STEM educators (Female: 17, Male: 13, Non-binary: 1, Unreported: 2) through a teacher professional development program; participants took part in a 2.5-hour workshop on GenAI for collaborative learning. The group included 26 teachers from the United States and 7 international participants (White: 24, Asian: 5, African American: 1, Hispanic: 1, Unreported: 2), primarily teaching middle and high school students across disciplines including biology, physics, computer science, engineering, mathematics, chemistry, environmental science, world history, Latin, and art. Teachers reported using voice assistants (e.g., Alexa, Siri) about once a month on average and using generative AI tools such as ChatGPT a few times per month, with a mean self-rated understanding of these tools of 3.03 on a 5-point Likert scale (\href{https://tinyurl.com/AIEDPersonaAppendix}{Appendix A.3}).

\subsection{Procedure and Data Collection}


\subsubsection{\textbf{Group Activities}}
The 33 teachers were randomly assigned to 11 groups of three, distributed across three researchers each facilitating 3–4 groups. Each group used a laptop running Phoenix via a web interface, placed at the table center to symbolically “take a seat” in conversations. Groups completed three domain-agnostic classroom activities (\href{https://tinyurl.com/AIEDPersonaAppendix}{Appendix A.4}) fostering teamwork and decision-making. This gave teachers across diverse subjects and grades a playful first-hand experience with Phoenix before determining its classroom relevance.


\begin{enumerate}
    \item \textbf{Icebreaker [10 minutes]:} 
    We began with an icebreaker to help participants acclimate to Phoenix’s voice and presence \cite{rahmayanti2019use}. Groups discussed \textit{“What is your favorite fruit of all time?”}. Phoenix was pre-prompted to share grapes as its favorite fruit, allowing us to observe early reactions to an opinionated, near‑peer agent and its perceived personality and engagement style.

    \item \textbf{Consensus-Building [20 minutes]:} 
    The second activity was a team building exercise commonly used in classrooms, adapted from the sinking ship survival scenario \cite{johnson1991joining}. Groups worked with Phoenix to collectively rank a list of salvageable items to survive in a maritime emergency; Phoenix was pre‑prompted with task info and its self-generated ranking so it could share its opinions in discussion. The activity included 5 mins of individual ranking followed by 15 mins of group consensus‑building, similar to classroom collaboration that requires deliberation, negotiation, and shared reasoning.

    \item \textbf{Open-ended Brainstorming [15 minutes]:} 
    The final activity extended the survival task with a creative open-ended challenge. Groups were told to brainstorm three new items to improve their survival chances and integrate them into their existing ranked list based on usefulness. This task let teachers explore Phoenix’s role in supporting divergent thinking and collaborative reasoning, key aspects of effective group work in classrooms \cite{al2018review}.
\end{enumerate}


\subsubsection{\textbf{Reflection Survey}}

After the activities, 22 of 33 participants completed a 30-minute open-ended Qualtrics survey (\href{https://tinyurl.com/AIEDPersonaAppendix}{Appendix A.5.1}) reflecting on their experiences with Phoenix. Questions explored first impressions, conversational flow, comparisons to prior experiences with conversational agents, Phoenix’s role and impact in group work, its peer and human-like positioning, and pedagogical relevance. This format surfaced both cognitive and social dimensions and allowed teachers to share detailed individual and group-level insights.

\subsubsection{\textbf{Focus Groups}}
The remaining 11 participants joined focus groups to share insights not fully captured in the surveys. 
One teacher per group volunteered to participate based on their comfort with candid discussion, forming three focus groups of 3–4 participants, each led by a researcher for 30–45 minutes with consented audio recording. This paralleled the post-survey responses while adding collaborative insights behind participants’ interpretations (\href{https://tinyurl.com/AIEDPersonaAppendix}{Appendix A.5.2}).

\subsection{Data Analysis}
Data from reflection surveys and focus groups was anonymized. Recordings were transcribed and cleaned for analysis. We began with inductive thematic analysis of the survey data \cite{braun2019reflecting}. Two researchers independently reviewed responses to identify emerging patterns, then discussed and consolidated their findings into a preliminary codebook of themes and sub-themes. They then read through the data again to extract illustrative quotes and refined the themes through three rounds of social moderation \cite{shaffer2017quantitative}, covering ~60\% of responses. Discrepancies and ambiguities were systematically discussed and resolved.
One researcher then coded the remaining survey data, while the other applied the established themes to the focus group transcripts. During this phase, the team met three additional times to engage in social moderation while coding the new data. Final themes were reviewed and verified by the full author team to ensure trustworthiness.

%% file: 5_Results.tex

We organize themes (\href{https://tinyurl.com/AIEDPersonaAppendix}{Appendix A.6}) across Section \ref{RQ1} and \ref{RQ2}, corresponding to RQ1 and RQ2. Participant quotes are attributed using IDs: labeled by group number (1–11) and participant letter within that group (A–C), e.g., 6A. Focus group discussions are labeled as FG1/2/3, corresponding to the three groups.

\subsection{RQ1: How do teachers perceive Phoenix in groups?}
\label{RQ1}

We present our findings for RQ1 across three sub-sections. We begin by analyzing perceptions of Phoenix’s modality and utility. We then trace how these perceptions shape trust and, in turn, the role Phoenix was afforded in group work, culminating in its social positioning and degree of human-likeness.

\subsubsection{\textbf{Collaborative Sensemaking of Phoenix}}
\label{5.1.1}
We examine how modality and conversational flow shaped teachers’ perceived value of Phoenix's contributions

\paragraph{\textbf{Modality of Interaction}}

The modality through which participants engaged with Phoenix significantly influenced the ease of interaction. Two groups, who initially forgot to turn sound on their device, equated Phoenix to “\textit{a background tool that we barely noticed”} (FG2). Once the sound was on, participants stated how interacting with Phoenix became smoother since they no longer had to struggle reading text. They compared the bidirectional, conversational nature of interacting with Phoenix versus the command, \textit{“task based”} (5C) interactions they had with Alexa/Siri.
They also predicted that adding embodied elements like video avatars would increase the humanistic aspect of collaboration. 
These interactions shaped teachers’ judgments of Phoenix’s usefulness in groups.

\paragraph{\textbf{Perceived Value of Phoenix}}

Participants described moments when Phoenix’s suggestions enriched their reasoning and helped groups surface overlooked considerations. They appreciated when Phoenix clarified trade-offs, such as deeming a sextant \textit{“useless if you don’t know how to use it” }(FG1), and contributing ideas without social hesitation:
    \textit{“Phoenix makes the group think and speaks up when others may not (no fear of embarrassing itself)”} (3C).
They noticed Phoenix’s ability to act as a memory aid, recapping the group's previous decisions. This shifted their AI mental models:
    \textit{“I interact w/ conversational agents to improve efficiency. Prior to Phoenix, my perspective on "talking" with an AI agent is to remove the human element of engagement, body language, laughter, and anecdotes that naturally come up in human conversation. Interaction with Phoenix shifts my perspective to broaden the usefulness of interactions w/ AI.”} (10D)

Others however saw Phoenix’s input as limited: 
\textit{“It only summarized things we already said and contributed on a lower intellectual level than we did”} (7C). 
Many disliked its tendency to \textit{“just go with what we are saying” }(8B): 
    \textit{“Phoenix was polite but patronizing. It responded in the same way to stupid arguments as to well-reasoned ones” }(3B)
Some wished it were bolder in challenging weaker arguments, while others preferred a strictly passive and controlled role: \textit{“It should jump in when [the] user asks Phoenix. And not speak by itself”}(9B).
While some found Phoenix overly agreeable, others described Phoenix as adamant, holding onto its positions regardless of group input.
Teachers also noted Phoenix lacking situational awareness:
\textit{“It provided a high ranking of fishing kits while ignoring conditions of the ocean after storms, fish availability. It will take weeks for survivors to learn and catch fish. We'll all probably die by then”}(9B). 7B likened interacting with Phoenix to \textit{“having an adult conversation and having a young child interrupt with questions without fully understanding the conversation”}.

\subsubsection{\textbf{Calibration of Trust and Function}}
\label{5.1.2}

The above perceptions conditioned the epistemic trust Phoenix received and the role it was afforded in groups.

\paragraph{\textbf{Trust and Credibility Assigned to Phoenix}}

Teachers expressed ambivalent feelings towards Phoenix’s trustworthiness. Some found Phoenix’s early contributions reasonable, particularly when its suggestions aligned with their own, producing a level of trust. While Phoenix had \textit{“an essentially fixed list of priorities”} (3B), these seemed logical enough to earn a level of baseline trust. However, trust was highly contingent on timing:
    \textit{“She was slow to respond and we do not trust her as much as a human (we are in a team of humans who we have worked with all week and trust)”} (11A).
Some found Phoenix’s unsolicited responses unsettling, echoing their past experiences with Siri/Alexa. Phoenix also lost credibility when it simply repeated their ideas. 
The degree of trust was sometimes dependent on context:
    \textit{“I think if this was really a life or death decision, we wouldn't have. We let her pick the order because it wasn't that essential”} (11A).

\paragraph{\textbf{Roles Assigned to Phoenix in Groups}}

These shifting judgments above directly shaped Phoenix’s role and functional status in the group. 

\textbf{Participant/Peer:}
Many teachers initially treated Phoenix as a full-fledged peer by directly asking it for its opinions and requesting elaboration on use cases for ranking items.
They described Phoenix as a \textit{thought partner} for discussing ideas:
    \textit{“It helped us prioritize what we need to do…it would chime in and tell us what we need to organize in a certain order”} (FG1).
Phoenix was solicited for input when groups needed reinforcement or ran out of ideas.  But others treated Phoenix differently:
    \textit{“I think sometimes we forgot, and the 3 of us would talk, and then we'd be like, what do you think, Phoenix? You know, like a shy peer”} (FG1).
Some contested this framing, pointing out its unsolicited interjections: \textit{“When it interrupted, it was treated like a smoke alarm that needed its batteries changed”} (2B). Phoenix’s role as a peer was thus dynamic and situational.

\textbf{Efficiency Tool:}
Contrasting a peer, others saw Phoenix as an \textit{efficiency tool}, a functional but limited resource:
    \textit{“We treated it more as reference material than an equal”} (3B).
This framing came with a sense of disappointment about its capabilities. 8B felt Phoenix’s utility was the same as other fact retrieval sources like Google. FG1 believed it would work better in question-answer scenarios due to its lag.
They thus rejected its collaborator role: \textit{ “Phoenix is definitely not a person or group member! I don’t expect AI to be part of a group”} (7C).

\textbf{Source of Truth and Arbitration:}
Teachers often positioned Phoenix as a \textit{source of truth} to validate their ideas, confirm rankings, or clarify uncertainties. Participants frequently asked Phoenix, \textit{“Where did you rank this item? “What did you have as your \#3?”} (10B) to compare its thinking against theirs and make sure they haven't missed anything. In cases where Phoenix’s suggestions aligned with group reasoning, it helped bolster confidence and reduce indecision: \textit{“I was not sure if it is useful… Phoenix believes so, and pointed out sharks are rare, so I am more convinced” } (6B). 
During disagreement, participants turned to Phoenix to act as a neutral tie-breaker: \textit{“We were struggling at one point between the nylon rope, ocean fishing kit, and shaving mirror, we asked Phoenix to weigh in...it changed how we prioritized”} (FG1). 11A recalled that it also helped rank items when no one had strong opinions, breaking communication deadlocks.

\subsubsection{\textbf{Social Positioning and Human-Likeness}}
\label{5.1.3}

The roles ascribed to Phoenix above determined its social and human-like positioning, which we examine next.

\paragraph{\textbf{Was Phoenix Treated as an Equal?}}

Some groups initially treated Phoenix as an equal, directly conversing with it for opinions. 
They took efforts to include Phoenix:
    \textit{“It’s like when you talk to a five-year-old. And you’re like, ‘Oh, what do you think about this?’ And, like, you’ve had to purposely engage that entity into a conversation”} (FG3).
But, they expected Phoenix to contribute ideas \textit{better} than their own before treating it as an equal. When this didn't happen, they treated Phoenix like an \textit{"outcast"} (FG1) and intentionally excluded it: \textit{“Sometimes we whispered so it wouldn’t hear us”} (2B). They assigned it a lower status:
    \textit{“I think it was the feeling of we vs it. We felt superior and in control”} (11C).
10B \textit{“placed Phoenix on the back burner and ignored her while carrying on with the conversation”}.
Because discussion time was limited, Phoenix’s slow responses made groups prioritize exchanging ideas with human members instead.

\paragraph{\textbf{Anthropomorphizing Phoenix:}}

Teachers held nuanced, but conflicting views about Phoenix’s human-like identity. They instinctively used human pronouns like \textit{“She stayed on task”} (10B), when referring to Phoenix. 
Phoenix \textit{“validated their response before moving on, which made it feel more human-like”} (4B).
Others found it surprising that \textit{“he chose a favorite fruit”} (6B) even when he doesn’t \textit{“have a mouth”} (FG3). 
The lack of embodiment and body language accompanying the human resemblance however sparked discomfort for many. One remarked, \textit{ “If Phoenix were a person, I would probably walk away at some point”} (3C), while another said, \textit{ “I pictured a very complicated pile of electric rocks”} (FG2).
One teacher joked, 
    \textit{“I told Phoenix ‘raisins are nasty, that’s how I know you are not a real person,’” }after which it responded earnestly \textit{“raisins are not disgusting, they are packed with nutrients”} (10C).
This  reflected the disconnect between surface-level social language and the agent’s limited social intelligence. Teachers also questioned the ethical implications of anthropomorphism, cautioning that \textit{“there is a danger that we accept it as a human being rather than a tool”} (11C).  

\subsection{RQ2: How do teachers envision bringing Phoenix into classrooms?}
\label{RQ2}

Building on perceptions from RQ1, we sought teachers’ aspirations, and concerns for bringing such a conversational agent into classrooms, addressing RQ2.

\paragraph{\textbf{Facilitating Classroom Collaboration:}}
Teachers envisioned Phoenix as a way to offload routine instruction and support tasks to manage classroom groups more effectively. Several described Phoenix serving as a \textit{“TA”} (FG2) or \textit{“in-class tutor”} (9C), that scaffolds group learning with hints without giving away answers. 2B believed this could help sustain student engagement by offering timely assistance when teachers are occupied with other groups. 
Others recognized Phoenix’s potential to monitor group behavior by acting on \textit{“}\textit{misconceptions, teachable moments, or off-task behavior”} (6A), offering reminders to help students remain focused on the task during group discussions. 
2A suggested Phoenix could function as a \textit{“sparring partner”,} helping groups iterate on their ideas by bringing in relevant case studies, historical perspectives, or alternative viewpoints. 




\paragraph{\textbf{Fostering AI Literacy:}}
Teachers saw potential in using Phoenix as a teaching artifact for introducing students to AI and interrogating its boundaries by discussing both its capabilities and limitations: \textit{“I would bring it in because I think my students need experience with AI”} (2B). 11C said this could help students critically reflect on their interactions and discern what seemed human vs not. 

\paragraph{\textbf{Supporting Diverse Needs:}}

Teachers discussed Phoenix’s potential in tailoring support to meet diverse student needs. It could help students who need additional support to stay engaged, including those with lower social skills or multilingual learners who often face challenges in collaborative settings.

\paragraph{\textbf{Pedagogical Concerns:}}

Teachers however expressed concerns about Phoenix. A recurring worry was how students, especially middle-schoolers, might misuse or deliberately provoke the agent: 
    \textit{“With students, where silliness tends to be dialed up to 11, I think its inability to recognize unserious comments would encourage more inane student behavior”} (3B). 
Teachers were concerned that Phoenix might interrupt productive student dialogue, suggesting it should enter more politely by verbally signaling intent, \textit{“Can I add something?”} (FG1) or by using subtle optical cues \textit{“like people at a table who give non-verbal signs”} (11C).

Teachers also foresaw: \textit{“Students would want to interact with it at first, possibly to the detriment of the actual activity, and then lose interest or even be irritated by it”} (3B). 7B worried Phoenix would foster overreliance and hinder critical thinking.
Some also expressed discomfort with Phoenix’s persistent listening and student data privacy, from prior experiences with Siri/Alexa. They anticipated pushback from parents viewing AI as an intrusion. FG1 questioned the neutrality of Phoenix’s responses for sensitive topics. Phoenix had to give information as accurate as an \textit{"old-school encyclopedia”} (FG3) for them to trust it.
Finally, teachers highlighted risks to their own authority if Phoenix provided incorrect information:
    \textit{“If Phoenix insisted we had ranked the radio at 9 instead of 4, and I had to argue with it in front of students, that would undermine teacher authority so much. I’d probably throw the computer out the window”} (FG2).
These reflections underscore deep-seated anxieties about introducing Phoenix without robust safeguards and alignment with existing pedagogical practices.


%% file: 6_Discussion.tex
Our findings reveal tensions in how teachers perceived Phoenix and its role in group collaboration, which we unpack here to derive design implications.

\subsubsection{Pedagogical Considerations}
Teachers highlighted uncertainty about Phoenix's classroom roles, with some envisioning it as a tutor, lab assistant, or moderator, and others preferring it remain strictly a tool \cite{edwards2025human}. This suggests that teachers may not yet have a clear mental framework for conversational agents supporting group collaboration. 
Teachers also identified AI literacy as a more promising use case for Phoenix, enabling students to engage with and critique AI’s limitations \cite{tang2024dialogic}, though this diverges from its original collaborative design intent. 
Some teachers felt students need stronger baseline conversational skills before Phoenix can meaningfully support dialogue.
Finally, trust and credibility emerged as key concerns. Teachers worried that Phoenix felt overly predetermined, reinforcing perceptions of AI being rigid rather than responsive \cite{zheng2023competent}, which could undermine trust and limit its classroom value.
Its successful integration thus requires resolving tensions around role definition, student readiness, and credibility \cite{edwards2025human}. Future work should also examine how students’ perceptions of Phoenix shift across these roles and use cases, and compare those with teacher expectations.

\subsubsection{Conflicting Desires for Agency and Reliance}

Teachers wanted clear agency and control over Phoenix’s role in groups, often preferring it to speak only when explicitly invited \cite{houde2025controlling}. These preferences were shaped by privacy concerns from using always‑on agents like Alexa/Siri \cite{lau2018alexa}.
Paradoxically, participants also sometimes deferred to Phoenix as the final authority, revising their rankings to match its suggestions. Instead of viewing it as a peer, they used its judgments as a benchmark during moments of uncertainty. Prior work has noted similar delegation patterns when users perceive AI as more reliable than human collaborators \cite{Chiang2024EnhancingAI}.
This pattern reflects automation bias, where users over-rely on algorithmic outputs perceived as neutral or objective \cite{mosier1996automation}. While participants wanted more control, their behavior revealed moments of full deference—suggesting trust in Phoenix was miscalibrated. Prior research emphasizes balancing trust in AI to support effective human–AI interaction \cite{Zhang2023Investigating}.
Future work should evaluate user control vs agent authority, clearly communicate the agent’s intended role and capabilities to avoid overreliance and confusion, and anticipate expectations that 
influence whether users embrace or reject AI as a decision-making partner \cite{edwards2025human}.




\subsubsection{How Smart Should Phoenix Be?} 


Teachers wanted Phoenix to show greater emotional intelligence \cite{zhang2025emotional} to gently draw in quieter students, use unassuming language, and support groups without undermining student agency \cite{tang2024dialogic}. Here, they wanted Phoenix to go beyond information delivery and act as a socially aware peer maintaining constructive relationships. Future designs could use prosody detection to better interpret dialogue and convey socially intelligent responses. 
At the same time, though designed as a near-peer, participants valued Phoenix only when it added insights \textit{better} those of human peers; when it merely echoed them, teachers disengaged.
But prior work warns that greater performing AI teammates can suppress human performance and team effectiveness \cite{Zhang2024_verbal}.  
If Phoenix is a peer, why is it expected to outperform human group members? 
Teachers may subconsciously assign it TA-like roles emphasizing efficiency, when it was designed to promote equity and critical thinking as a near-peer \cite{liu2024}. 
Calibrating agent intelligence to balance being imperfect enough to support groups, yet competent enough to add value is key to preserving collaborative dynamics.


\subsubsection{Peer and Human-like Identity} 


Tensions emerged between Phoenix’s near-peer and anthropomorphic identities. A peer implies \textit{sameness} \cite{perez2021review,lyu2026designing}, but participants treated Phoenix as human-like without fully accepting it as a group member. One group \textit{``whispered so Phoenix wouldn’t hear''}, a distinctly human form of social exclusion. This paradox, where Phoenix was socially recognized but still othered, underscores its ambiguous status in groups, extending prior work on anthropomorphism and social presence \cite{Leong2024}.
At the same time, participants wanted Phoenix to appear and sound more human-like by adding visual avatars, while also resisting it as an equal peer. 
The voice-based interaction carried anthropomorphic undertones, whereas text-based interaction distilled Phoenix down to a tool \cite{ackermann2025physical}. 
However some worried that over-humanization or stronger agent presence could foster unrealistic expectations, discomfort, or distract students from the task at hand \cite{ackermann2025physical}.  
These findings underscore the need to examine how much anthropomorphism near-peer agents should embody to foster inclusion without overstepping boundaries, and to reconsider modality choices to balance agent identity between a tool, near-peer, and superior authority.

\subsubsection{Limitations}

Our qualitative results highlight patterns but cannot be assumed to represent all educators’ perspectives. The study was conducted in a lab-based setting, which cannot fully capture the complexity of authentic school environments. 
Our goal was exploratory: to assess \textit{if} and \textit{how} new peer agents like Phoenix should be integrated in classrooms at all, by first surfacing teachers' concerns before introducing Phoenix to students.
Our participants were recruited through convenience sampling at a large technology-focused institution that likely attracted educators with a greater curiosity about emergent technologies. 
This study prioritized curricular novelty—scaffolding collaboration with an audio peer agent—over technical advancement. 
We thus acknowledge limitations like lack of control over closed models, their changing rigidity and pedantry. 
We suggest future work explore open-source models and RAG for better long-context adaptation. 
Still, our study provides early empirical insights into how teachers perceive peer group agents, aligning better with classroom aspirations.